\documentclass[a4paper,fleqn]{cas-dc}

\usepackage[numbers]{natbib}
\let\printorcid\relax 
\usepackage[flushleft]{threeparttable}
\usepackage{hyperref}
\usepackage[nameinlink,capitalize]{cleveref}
\usepackage{microtype}
\usepackage{graphicx}
\usepackage{caption}
\usepackage[caption=false]{subfig}
\def\tsc#1{\csdef{#1}{\textsc{\lowercase{#1}}\xspace}}
\tsc{WGM}
\tsc{QE}
\tsc{EP}
\tsc{PMS}
\tsc{BEC}
\tsc{DE}

\begin{document}

\captionsetup[figure]{labelfont={bf},labelformat={default},labelsep=period,name={Fig.}}
\let\WriteBookmarks\relax
\def\floatpagepagefraction{1}
\def\textpagefraction{.001}
\shorttitle{Automatic Detection of Abnormal EEG Signals.}
\shortauthors{Albaqami Hezam et~al.}

\title [mode = title]{Automatic~detection~of~abnormal~EEG~signals~using~wavelet~feature~extraction~and~gradient~boosting~decision~tree}

\author[1]{Hezam~Albaqami}
\cormark[1]

\ead{hezam.albaqami@research.uwa.edu.au}

\author[1]{Ghulam~Mubashar~Hassan}

\author[2]{Abdulhamit~Subasi}

\author%
[1]{Amitava~Datta}

\address[1]{Department of Computer Science and Software Engineering, The University of Western Australia, Australia}
\address[2]{Department of Computer Science, Effat University, Saudi Arabia}

\cortext[cor1]{Corresponding author at: Department of Computer Science and Software Engineering, The University of Western Australia, 35 Stirling Hwy, Crawley WA 6009, Australia.  Tel.: +61 8 6488 2281.}

\begin{abstract}
Electroencephalography is frequently used for diagnostic evaluation of various brain-related disorders due to its excellent resolution, non-invasive nature and low cost. However, manual analysis of EEG signals could be strenuous and a time-consuming process for experts. It requires long training time for physicians to develop expertise in it and additionally experts have low inter-rater agreement (IRA) among themselves. Therefore, many Computer Aided Diagnostic (CAD) based studies have considered the automation of interpreting EEG signals to alleviate the workload and support the final diagnosis. In this paper, we present an automatic binary classification framework for brain signals in multi-channel EEG recordings. We propose to use Wavelet Packet Decomposition (WPD) techniques to decompose the EEG signals into frequency sub-bands and extract a set of statistical features from each of the selected coefficients. Moreover, we propose a novel method to reduce the dimension of the feature space without compromising the quality of the extracted features. The extracted features are classified using different Gradient Boosting Decision Tree (GBDT) based classification frameworks, which are CatBoost, XGBoost and LightGBM. We used Temple University Hospital EEG Abnormal Corpus V2.0.0 to test our proposed technique. We found that CatBoost classifier achieves the binary classification accuracy of 87.68\%, and outperforms state-of-the-art techniques on the same dataset by more than 1\% in accuracy and more than 3\% in sensitivity. The obtained results in this research provide important insights into the usefulness of WPD feature extraction and GBDT classifiers for EEG classification.
\end{abstract}


\maketitle
\section{Introduction}\label{sec:introduction}
The neural activity of human brain begins at an early stage of prenatal development. The brain electrical signals represent the brain functions and the status of the whole body throughout life~\cite{Sanei2008}. In 1875, Richard Caton produced the first recorded brain activity in the form of an electrical signal by placing two electrodes on the scalp. Since then, the term EEG stands for brain electrical neural activity~\cite{Sanei2008}. Electroencephalography (EEG) is an electrophysiology monitoring method of recording the brain electrical activities. It measures the brain activities by placing a set of sensors (electrodes) on the scalp. It is a primary clinical tool used to diagnose epilepsy and stroke~\cite{Obeid2018}. With advances in technology, many applications have been developed that allow health clinics to use EEG for diagnostic evaluation of various brain disorders including, but not limited to, Alzheimer's, sleep disorder and head-related trauma~\cite{Obeid2018}.

The signals recorded from the scalp are digitized and presented in a waveform. EEG specialists examine the waveform and generate a diagnostic report stating the status of the individual. Usually, the diagnostic first step is to decide whether the recorded brain signals show abnormal behavior or not. Then the brain disorder related to this abnormality can be further investigated and the required medication is provided. Clinical use of EEG is increasing rapidly due to its ease of use as a non-invasive procedure. Moreover, EEG is a comparatively inexpensive diagnostic tool which makes it a popular choice among physicians~\cite{Nannen:Thesis:2017}. However, manual interpretation of EEGs is a time consuming and overwhelming process. Besides, it requires a tremendous amount of training time for physicians to be experts in it. Moreover, EEG signal analysis has a low inter-rater agreement (IRA) even among highly certified specialists~\cite{Obeid2018}.
Therefore, there is a need to develop automated procedures to interpret EEG in real-time to preserve physicians' time and aid them in making accurate decisions~\cite{Obeid2018,Schirrmeister2018}.

Recently, automatic analysis of EEG has gained attention due to the robustness of machine learning algorithms, advances in and low cost of high-performance computing, and the availability of big data~\cite{Roy2019}. The applications are mostly focused on the diagnosis of epileptic seizures~\cite{shoeb2010application,asif2019seizurenet, 9055195}, depression~\cite{Acharya2018depression}, brain trauma~\cite{Naunheim2010trauma} Parkinson's~\cite{Oh2018parkinsons}, Alzheimer's~\cite{kabbara2018reduced}, and general EEG pathology~\cite{Nannen:Thesis:2017,Schirrmeister2018,yildirim2018deep,roy2019chrononet,Alhussein2019, Amin2019,Gemein2020,van2019detecting}. The variety of brain health conditions that can be addressed by EEG indicates that there is a high potential in interpreting EEGs automatically as well as understanding the manifestation of different diseases using EEGs~\cite{Obeid2018}.

Automatic classification and interpretation of EEGs have been investigated by many methods such as time-frequency analysis~\cite{gotman1999automatic,sartoretto1999automatic}, nonlinear techniques~\cite{schindler2001using} and expert systems~\cite{Khamis, subasi2010comparison}. Recently, different machine learning techniques are proposed for the same task such as Support Vector Machine (SVM)~\cite{alotaiby2014eeg}, k-nearest neighbors~\cite{Lopez}, Random Forest (RF)~\cite{MursalinRandom}, linear discriminant analysis~\cite{subasi2010comparison}, logistic regression~\cite{Alkan2005logistic} and neural networks~\cite{ramgopal2014seizure,singh2019usage,orhan2011eeg,VijayAnand2019,kocadagli2017classification, Gemein2020}.

Although the recent research trend is shifting towards deep learning based models where raw data is provided as inputs, still feature extraction based Artificial Intelligence (AI) techniques have high potential to perform well. Different feature extraction techniques are proposed over time such as time-domain, frequency-domain and time-frequency domain~\cite{Lopez,subasi2007eeg,Gemein2020}, and the extracted features are provided as inputs to AI based classification models. It is observed from the literature that both feature extraction methods as well as classification models play critical roles in the overall performance of a system.

 \sloppy
In the context of feature extraction, time-frequency analysis has been reported to perform well in capturing different behaviors of non-stationary signals~\cite{bigan1998recursive,Subasi2019}. Moreover, Wavelet Transform (WT) algorithms are successful tools for extracting robust features because of their suitability for capturing transient events such as spikes and sharp waves~\cite{Subasi2019,bigan1998recursive}. Continuous Wavelet Transforms (CWT),~Discrete Wavelet Transforms (DWT) and WPD are different variants of WT which have been reported to extract features from EEGs and are used in various EEG processing such as Brain-Computer Interface BCI~\cite{kevric2017comparison, ting2008eeg}, and biomedical signal processing~\cite{singh2019usage,subasi2007eeg,bigan1998recursive,subasi2010comparison,VijayAnand2019,kocadagli2017classification,volschenk, acharya2011automatic}. However, selection of important feature from all the extracted features is still open ended problem.

Subasi et al.~\cite{subasi2019comparison} selected a set of features from WPD extracted features and combined them with Random Forest (RF) classifier. The study mentioned that the selected WPD features are superior to DWT extracted features for the task of classifying focal and non-focal EEGs. Acharya et al.~\cite{RajendraAcharya2012} used WPD and selected set of features from EEGs for epileptic activity classification. Kutlu et al.~\cite{kutlu2012feature} proved in an experimental study that features extracted by WPD are highly efficient for the classification of Electrocardiogram signals (ECGs) for heart problems. Moreover, WPD has also been used successfully in general pattern recognition problems such as in near-infrared spectra~\cite{walczak1996application} and faulty bearing diagnosis in machine monitoring~\cite{berredjem2018bearing}.  

In the context of classification methods, GBDT is a widely popular choice for classification and regression problems. GBDT is a powerful machine learning algorithm which has different practical implementations including but not limited to Catboost~\cite{NIPS2018_7898}, XGBoost~\cite{Xgboost} and LightGBM~\cite{lightGBM}. These algorithms have been applied successfully in many applications, including EEG signal processing and classification. They have resulted in providing state-of-the-art results in many machine learning tasks~\cite{Xgboost} and are widely used in both industry and academia. For example, in~\cite{multi-person}, XGboost is employed for multi-person brain activity recognition based on EEGs. The experimental results achieved state-of-the-art results at the time of its publication. Indeed, XGBoost is used by most of the teams in Kaggle competition that was held in 2016 to predict seizures in long-term human intracranial EEG recordings~\cite{awy210}. Overall, GBDT is a popular and successful choice in many popular solutions of machine learning competitions for various tasks, including temporal data~\cite{Xgboost}.

In this study, we focused on the publicly available EEG TUH Abnormal EEG Corpus dataset~\cite{Nannen:Thesis:2017}. It is a subset of the TUH EEG Corpus~\cite{obeid2016temple} and is the most comprehensive and regularly updated open-source EEG dataset available for researchers. To the best of our knowledge, there are only eight research studies for EEG pathology detection and all of them are using TUH Abnormal EEG Corpus~\cite{Lopez} except Leeuwen et al.~\cite{van2019detecting} which used different dataset. Two of the studies considered classification using handcrafted features~\cite{Lopez,Gemein2020} while rest of the studies proposed classification models based on end-to-end deep learning models. Lopez et al.~\cite{Lopez} proposed a two-dimensional deep Convolutional Neural Network (CNN) and Multilayer perceptron (MLP) that uses precomputed band power-based features using Cepstral coefficients to detect abnormality in EEGs. The result of their study reported an accuracy of 78.8\%. Lukas et al.~\cite{Gemein2020} proposed a feature-based technique comparing two approaches: feature-based machine learning frameworks and end-to-end deep learning methods for EEG pathology detection. In the feature-based approach, the study proposed to extract a vast collection of EEG features using Discrete Fourier Transform (DFT), CWT, DWT and connectivity between electrodes using Hilbert Transform (HT). Various machine learning models are used for classification of extracted features such as Support Vector Machine (SVM), RF, Auto-Sklearn Classifier (ASC) and Riemannian-Geometry-based (RG) classifier with SVM. While in the deep learning approach, the study proposed to utilize different neural network architectures such as Convolutional Neural Networks (CNNs) and Temporal convolutional network (TCN) to do the same task. Different deep neural network architectures applied to the task of EEG binary classification are Braindecode Deep4 CNN (BD-Deep4)~\cite{Schirrmeister2018}, Braindeocde Shallow CNN (BD-Shallow)~\cite{Schirrmeister2018}, Braindecode EEG neural network BD-EEGNet~\cite{van2019detecting}, and an EEG-optimized TCN model (BD-TCN). The best-reported result for feature-based models achieved the accuracy of 85.9\% by RG classifier and the best-reported result in the deep learning approach achieved the accuracy of 86.2\% by BD-TCN~\cite{Gemein2020}. 
 
 Rest of the published studies only considered end-to-end deep learning models. Schirrmeister et al. proposed deep and shallow CNN models to detect irregular EEGs which provided an accuracy of 85.4\%~\cite{Schirrmeister2018}. Similarly, Yildirim et al. proposed a complete end-to-end deep CNN model with 23 layers for binary classification of EEGs~\cite{yildirim2018deep}. The study reported an accuracy of 79.34\%. Roy et al.~\cite{roy2019chrononet} proposed a deep gated recurrent neural network (ChronoNet) model to detect abnormality in EEGs. The reported result had an accuracy of 86.57\%. Amin et al.~\cite{Amin2019} deployed a popular CNN model ALexNet with SVM that was pre-trained on EEG data to detect anomalous events in EEGs. The study reported an accuracy of 87.32\%. Alhussien et al.~\cite{Alhussein2019} proposed a pre-trained deep CNN model AlexNet with MLP to detect abnormal events in EEGs which achieved an accuracy of 89.13\%. Both proposed networks in~\cite{Amin2019, Alhussein2019} were pre-trained on 10,000 normal EEG recordings (for which the details were not available).

From the analysis of the existing techniques, we observed that the error rate in classification is still high in most of the studies using this dataset and novel technique is required to improve the classification performance. It has been reported in the literature that the sensitivity of 95\% is the minimum requirement for an application to be clinically accepted~\cite{golmohammadi2019automatic}.
\begin{figure*}
	\includegraphics[width=\linewidth]{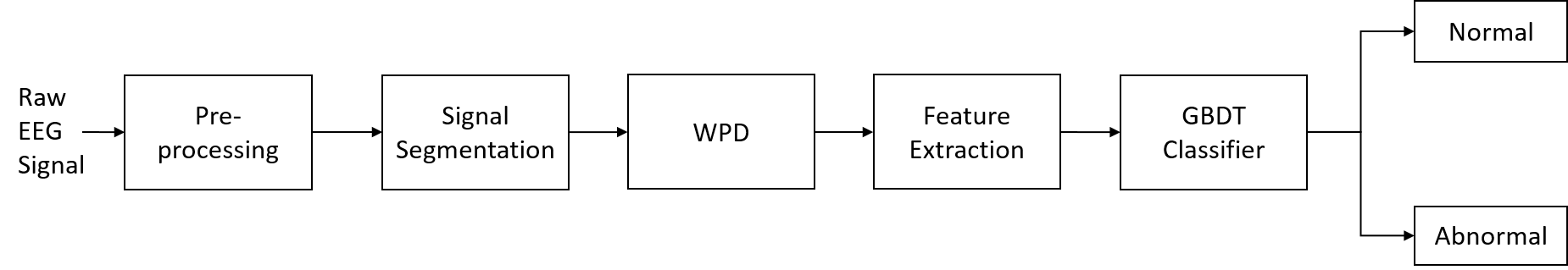}
	\caption {Proposed architecture for binary classification of EEGs.}
	\label{fig:overall}
\end{figure*}

In this research, we are proposing a novel technique for the problem of detecting abnormal brain signals in EEGs which extracts specific features from the EEG signals. We used WPD for feature extraction and selected a specific set of important features which play a critical role in improving the classification of EEGs. Our novel method proposes to aggregate the extracted features that helped in reducing the dimension of the feature space without compromising the quality of features. Three different GBDT classifiers were used to classify EEGs based on the extracted features in aggregated feature space. We evaluated the performance for the proposed technique on TUH Abnormal EEG Corpus~\cite{Nannen:Thesis:2017} which include signals of various brain-related disorders such as Alzheimer's, epilepsy, and stroke~\cite{Gemein2020}. The results obtained by our proposed technique were found to be better than the existing state of the art results. 

This article is organized as follows:~\Cref{sec2} contains the details of TUH Abnormal EEG Corpus dataset, the proposed technique in detail, and the performance evaluation metrics. ~\Cref{sec3} presents and analyses the results while~\Cref{sec4} concludes the study with future recommendations.

\section{Material and Methods}\label{sec2}
In this study, we propose to use WPD for feature extraction from EEGs and use GBDT classifiers to classify abnormal EEGs from normal ones. We test our proposed technique using publicly available dataset TUH Abnormal EEG Corpus~\cite{Nannen:Thesis:2017}.~\cref{fig:overall} shows the overall architecture of the proposed technique and steps involved in the binary classification of EEGs.

\subsection{Data}\label{data}
\sloppy \sloppy \sloppy \sloppy
Temple University Hospital EEG Corpus (TUH EEG) \cite{obeid2016temple} is the largest available open-source EEG database. It contains more than 25,000 EEG records for more than 14,000 subjects.The recent release of this corpus has many subsets focusing on different research problems. In this research, we used TUH EEG Abnormal Corpus V2.0.0 \cite{Nannen:Thesis:2017}. It comprises more than 2000 EEG sessions with at least 15 minutes duration each. 1521 recordings were labelled as normal and 1472 recordings were labelled as abnormal. It contains recordings for both genders with different ages that range from infants to seniors. The dataset is divided into training and evaluation subsets. The training subset consists of 2717 EEG recordings, (1371 normal/1346 abnormal) while the testing subset consists of 276 EEG recordings (150 normal/126 abnormal) as shown in \cref{tab:TUH abnormal dataset}. There is no overlapping between subjects in the training and the testing datasets~\cite{Nannen:Thesis:2017}.

\begin{table}[width=.9\linewidth,cols=4,pos=h]
\caption{TUH EEG Abnormal Corpus V2.0.0.}\label{tab:TUH abnormal dataset}
\begin{tabular*}{\tblwidth}{@{} LLLL@{} }
\toprule
 & Training & Testing \\
\midrule
Normal & 1371 & 150  \\
Abnormal & 1346 & 126  \\
Total & 2717 & 276  \\
\bottomrule
\end{tabular*}
\end{table}

\begin{figure}
    \centering
  	\includegraphics[width=0.8\linewidth]{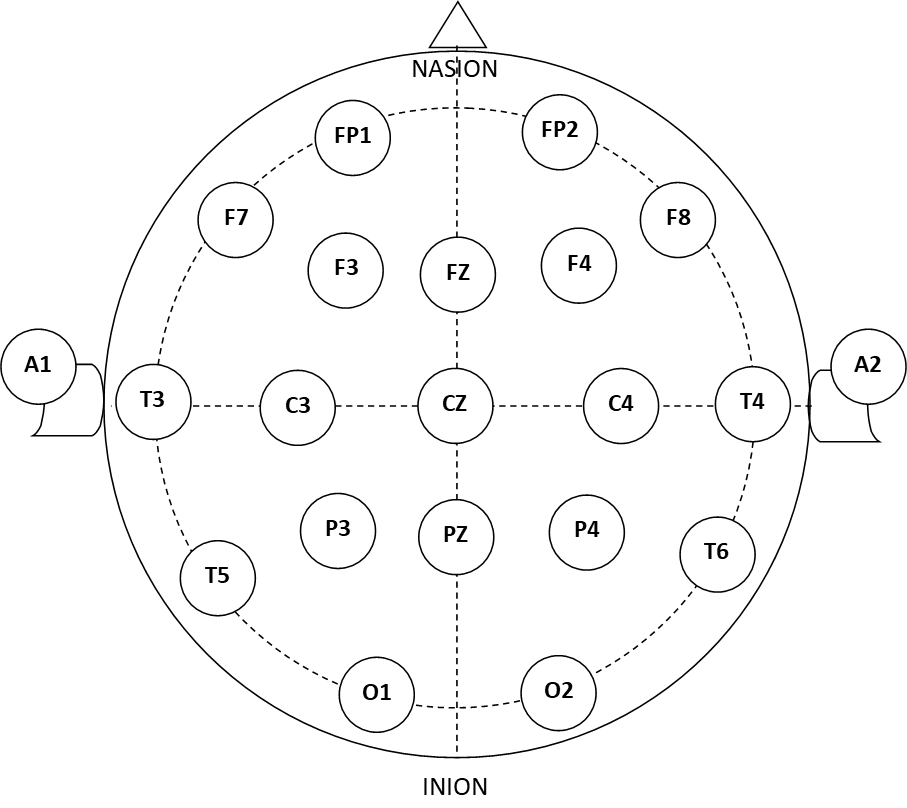}
  	\caption {The international 10/20 system of electrodes placement with even-numbered electrodes (2,4,6,8) and odd-numbered electrodes (1,3,5,7) on the right and left side of the scalp, respectively, while zeros (z) electrodes on the middle of the scalp.}
  	\label{fig:1020map}
\end{figure}

\subsection{Preprocessing}
\label{Preprocessing}
\begin{figure*}
    \centering
  	\includegraphics[width=\linewidth]{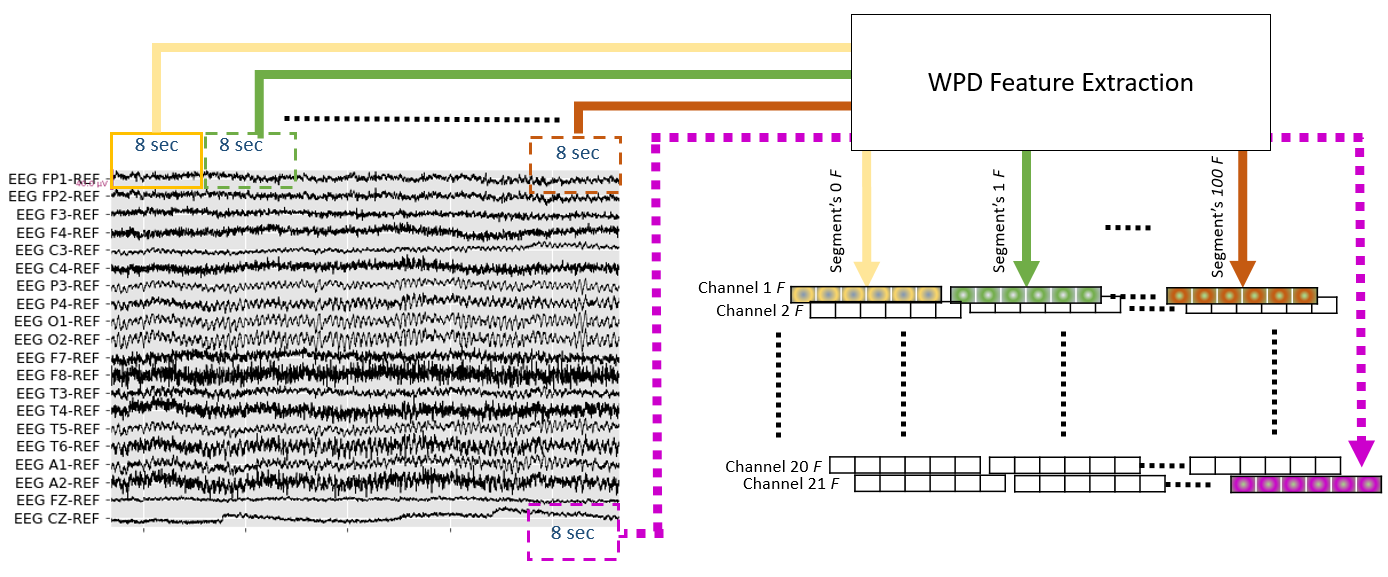}
  	\caption {EEGs segmented channel-wise into 8 seconds long segment. Features are extracted using WPD and aggregated into two-dimensional array. 
  	}
  	\label{fig:segmentation}
\end{figure*}

TUH EEG Abnormal Corpus dataset is based on 10/20 electrode configuration and collected using different devices at different times~\cite{obeid2016temple}. To maintain the consistency of the data, we ensured that the data of 21 EEG channels is only selected as we encountered some data with more than 21 channels. The selected 21 channels are presented in \cref{fig:1020map}, while data from extra channels was removed. The sampling rate varied between 250 Hz and 500 Hz; therefore, we re-sampled all the recordings at lowest frequency of 250 Hz. Afterwards, we segmented each signal (channel-wise) into 8 seconds non-overlapping parts as shown in \cref{fig:segmentation} which resulted in 2000 data points for each segmented part. 

\subsection{Wavelet Transform}\label{WaveletTransform}
        Since the 1980s, Wavelet Transform (WT) is used to de-noise and extract features from signals as it plays a major role in the analysis and classification of non-stationary signals similar to EEG having a high resolution in both time and frequency domain~\cite{faust2015wavelet,Subasi2019}. WT is defined as a spectral estimation technique that can represent any function as an infinite series of wavelets~\cite{vetterli1992wavelets}. It can capture the detailed alteration within the EEGs including abrupt changes. WT is broadly classified into Continues Wavelet Transforms (CWT) and Discrete Wavelet Transform (DWT). CWT of signal $x(t) $ is described as:
        
\begin{equation}
W_x (a,b) = \frac{1} {|a|^{1/2}}\int_{-\infty}^\infty x(t)
\overline{\psi} \left(\frac{t-b}{a}\right) dt\; \end{equation}
where $\psi(t)$ is the continuous mother wavelet and its complex conjugate is represented by $\overline{\psi(t)}$, $a$ is the dilation parameter, and $b$ is the translation parameter. $\psi(t)$ gets scaled and translated by factors $a$ and $b$, respectively, and the correlation coefficient is computed for the shifted interval in the $x$-axis. The process can be repeated for various scaling factors on the $y$-axis. 

\begin{figure}
	\centering
	\includegraphics[width=\linewidth]{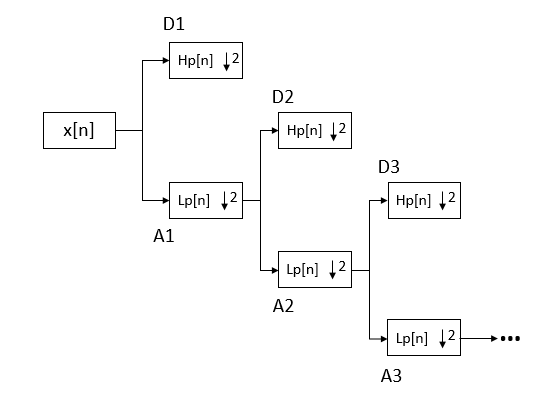}
	\caption {The DWT structure where the signal splits into high and low-frequency at each level but only lower frequency components are further split in the next level.}
	\label{fig:DWT_decomposition}
\end{figure}

DWT is another implementation of WT which covers the redundancy in CWT by decomposing a given signal into a mutually orthogonal set of wavelets. In DWT, both the $a$ and $b$ factors are limited to only discrete values. DWT decomposes a signal into approximation and detail coefficients using filter banks. The approximation coefficient is the output of the low-pass filter, and the detail coefficient is the output of the high-pass filter~\cite{meyer1992wavelets}. The approximation coefficient can be further decomposed into a lower level of approximation and detail coefficients. \cref{fig:DWT_decomposition} illustrates the process of DWT~\cite{Shoeb2004}. 

\begin{figure}
	\centering
	\includegraphics[width=\linewidth]{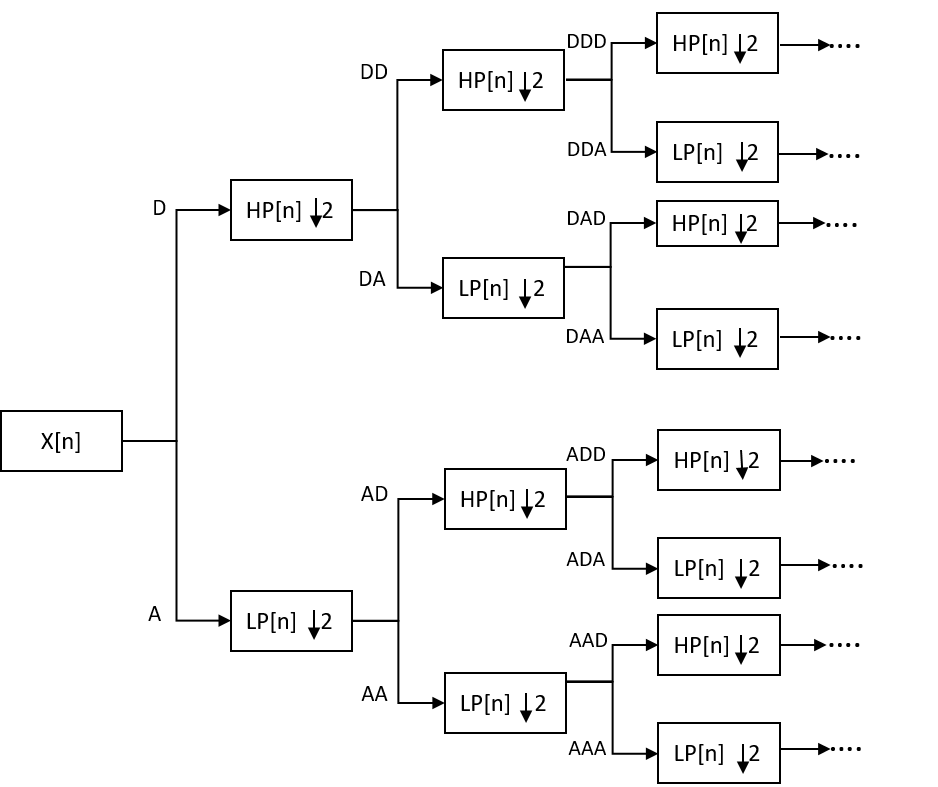}
	\caption {The structure of three scale level of WPD where the signal splits into high and low frequency components at each level.}
	\label{fig:PWD_decomposition_tree}
\end{figure}

WPD is also known as the Optimal Subband Tree Structure (SB-TS), is an extended version of DWT. It covers the deficiency of DWT, which only splits the low-frequency components of the signal. WPD decomposes both low as well as high-frequency components which leads to a complete wavelet binary tree as shown in \cref{fig:PWD_decomposition_tree}. Thus, WPD provides better frequency resolution, and we selected it to decompose the EEG signals into eight levels of decomposition as mentioned in Figure \ref{fig:newFeatures}. 

\subsection{Gradient Boosting Decision Tree classifiers}
\label{GBDT}
As mentioned earlier, GBDT is a popular supervised machine learning algorithm which is regularly used in automated biomedical solutions. The three recently introduced variants  are Catboost, XGBoost and LightBM. The hyper-parameters tuned for the three models are presented in \cref{tab:hyperparameters}, and their short details are described below. A more detailed description of the classifiers can be found in the mentioned studies.

\subsubsection{XGBoost}
\label{GBDT1} XGBoost stands for Extreme Gradient Boosting introduced by Yanki in 2016~\cite{Xgboost}. It is a scalable end-to-end tree boosting system that generates multiple weak learners that are trained consecutively one after another. Every new learner corrects the mistakes made by the preceding one to end up in a final optimal model. Owing to its simplicity and efficiency, XGBoost is popularly used to solve many data science tasks, including EEG signal processing~\cite{Xgboost,awy210,zhang2018converting}.
\begin{table}[width=.9\linewidth,cols=4,pos=h]
\caption{Hyper-parameters for the classifiers.}\label{tab:hyperparameters}
\begin{tabular*}{\tblwidth}{@{} LLLL@{} }
\toprule
Hyperparameter & Catboost & XGboost & LightGBM\\
\midrule
learning\_rate & 0.02 & 0.0156 & 0.0182 \\
max\_depth  & 5 & 8 & 10 \\
n\_estimators & 1500 & 300 & 250 \\

\bottomrule
\end{tabular*}
\end{table}

\subsubsection{LightGBM}
\label{GBDT2}
In 2017, Microsoft introduced a different type of GBDT algorithm known as Light Gradient Boosting Machine (LightGBM) \cite{lightGBM}.~It uses Gradient-based One-Side Sampling (GOSS) to filter out the data instances for finding the best split value. Moreover, it uses Exclusive Feature Bundling (EFB) technique that reduces the feature space complexity by taking advantage of the feature space sparsity to bundle together mutually exclusive features, e.g., it never holds a value of zero simultaneously. 

\subsubsection{CatBoost}\label{GBDT3}
Categorical Boosting (CatBoost) is one of the most recent GBDT algorithms which was introduced by Yandex in 2018~\cite{NIPS2018_7898}. It is capable of handling categorical data. It has two critical algorithms, ordered boosting, which calculates leaf values during the selection of the tree structure to reduce over-fitting, and an innovative algorithm for processing categorical features during training time instead of preprocessing. It is simple to implement and very powerful, flexible and portable. Moreover, CatBoost can work with various data types. It provides best-in-class accuracy with short learning time.

\subsection{Feature Extraction}\label{FeatureExtraction}
\begin{figure*}
	\centering
	\includegraphics[ width=\linewidth]{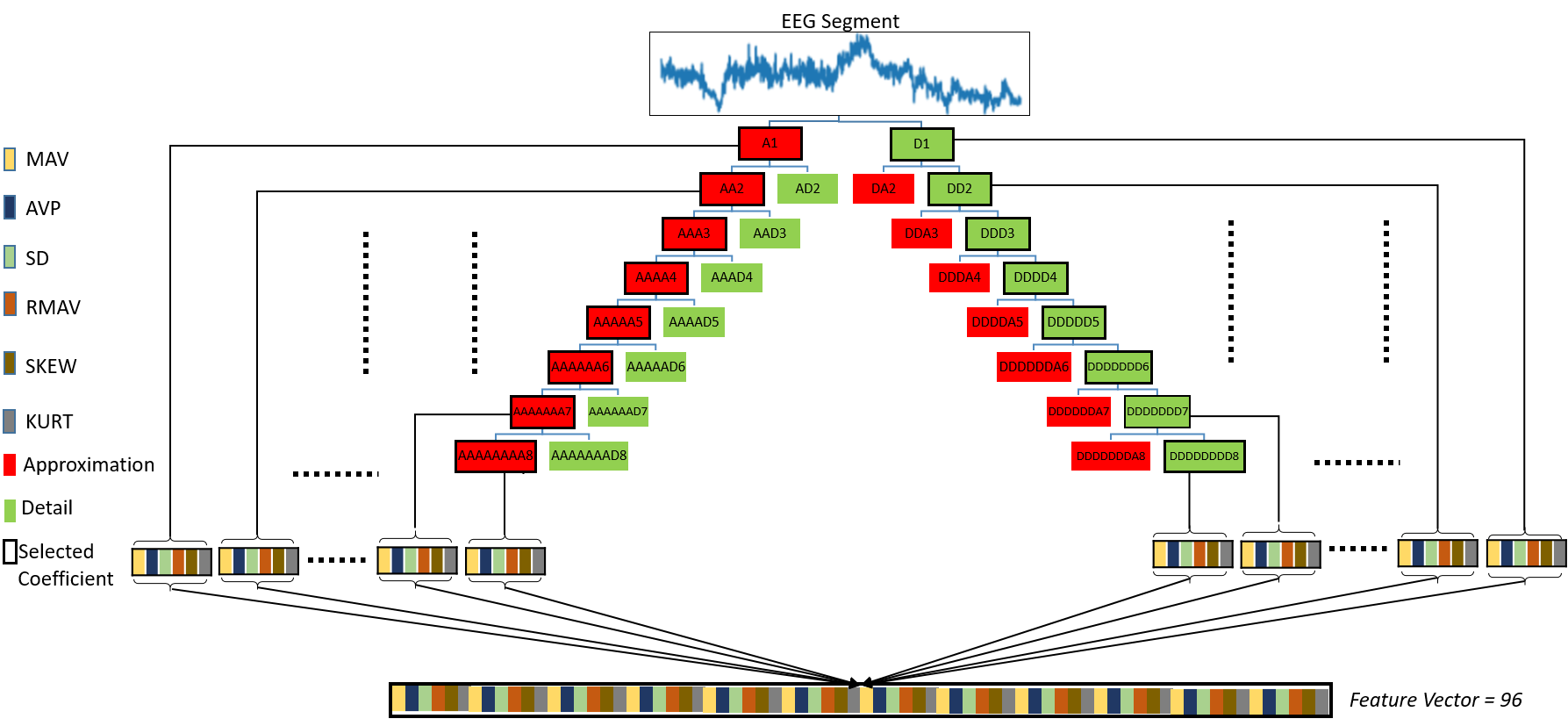}
	\caption {WPD feature extraction process which includes extraction of selected coefficients from a single EEG segment and computing 6 statistical features from each of the selected coefficients as explained in Section ~\ref{stfeatures}.}
	\label{fig:newFeatures}
\end{figure*}

We used WPD to decompose the EEG signal into different levels of frequency sub-bands. We chose the known mother-wavelet Daubechies 4 (db4)~\cite{daubechies1988orthonormal} for eight levels of decomposition to get the detail and approximation of the signal. Literature indicates that the wavelet filter db4 is the most suitable wavelet for abnormality detection in EEGs~\cite{faust2015wavelet, subasi2010comparison}. For the decomposed approximation coefficients, we selected its approximation nodes only. The same procedure is also applied to the detail coefficients as shown in \cref{fig:newFeatures}, which displays part of the decomposition and the selected coefficients.

Overall, we obtained 16 decomposed components: 8 approximation components and 8 descended from the detail components as presented by black borders on the left and right side in \cref{fig:newFeatures}. Although these decomposed signals can be used as feature vector for a machine learning algorithm, it has been observed that these decomposed signals are very sensitive to noise~\cite{shoeb2009application}. Therefore, we computed a set of statistical features from each of the decomposed levels as suggested in \cite{Subasi2019,subasi2019comparison}. \cref{fig:newFeatures} depicts the process of the computed features and the corresponding coefficients. The computed statistical features are:
\begin{enumerate}\label{stfeatures}
\item Mean absolute values of the coefficients in each sub-band (MAV).
\item Average power of the coefficients in each sub-band (AVP).
\item Standard deviation of the coefficients in each sub-band (SD). 
\item Ratio of the absolute mean values of adjacent sub-bands (RMAV). 
\item Skewness of the coefficients in each sub-band
(skew). 
\item Kurtosis of the coefficients in each sub-band
(Kurt). 
\end{enumerate}

After computing the features, the steps of feature normalization and aggregation are performed.

\begin{figure}
	\centering
	\subfloat[Sample feature vector with 20 features.]{\includegraphics[width=\linewidth]{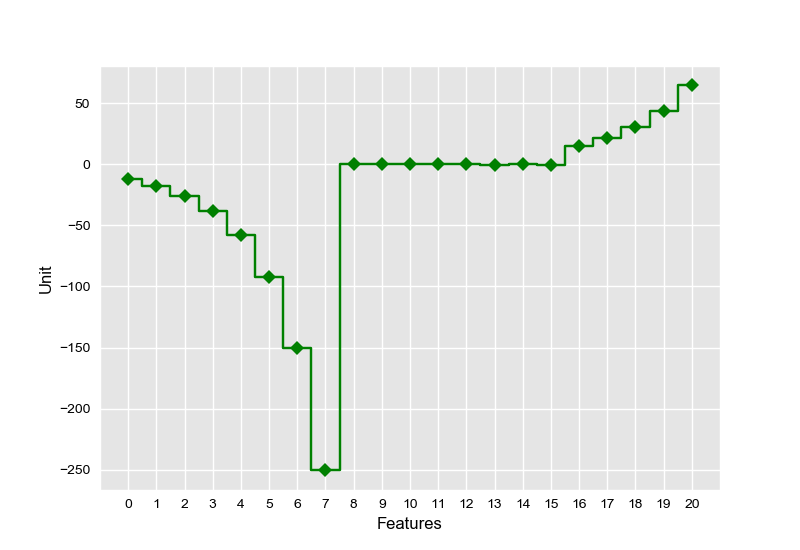}\label{fig:featurenormalization1}}
  \hfill
  \subfloat[Normalized sample feature vector with 20 features.]{\includegraphics[width=\linewidth]{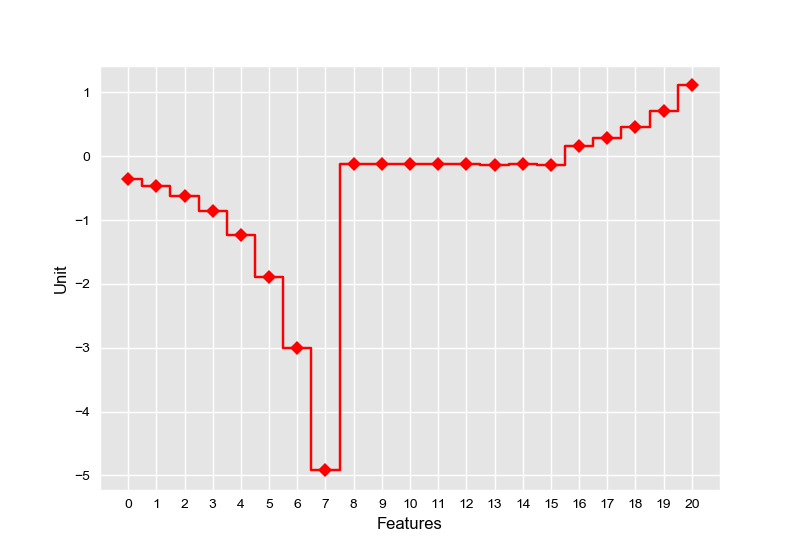}\label{fig:featurenormalization2}}
	\caption {Feature vector before and after normalization.}
\end{figure}
\subsubsection{Feature vector normalization}
\label{FeatureNormalization}

\begin{figure}
	\centering
	\includegraphics[height= 6cm,width=\linewidth]{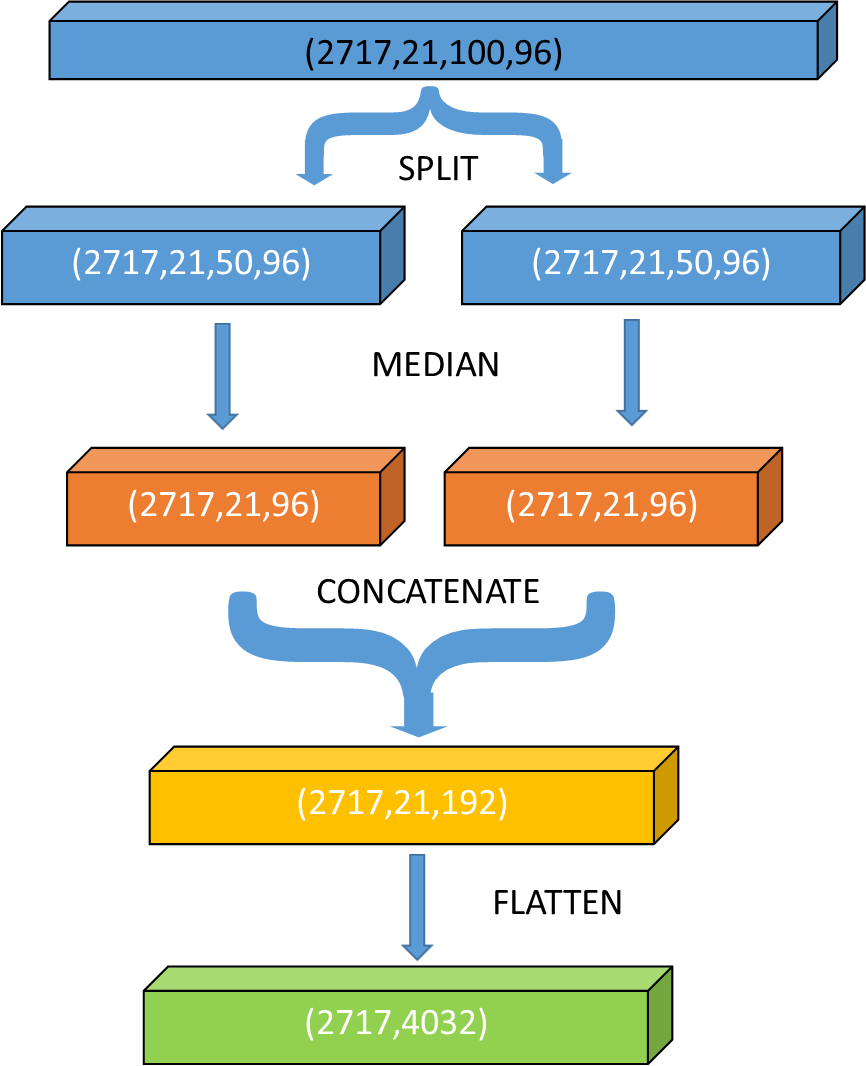}
	\caption {Feature Aggregation process.}
	\label{fig:split1}
\end{figure}
The computed statistical features have a wide range of values; therefore, we normalized each feature vector within a narrower range by using the standard scalar method~\cite{scikit-learnSTS}. The normalized computed feature vector had zero mean and standard deviation equal to 1. \cref{fig:featurenormalization1} and \cref{fig:featurenormalization2} show a feature vector of 20 features before and after normalization, respectively. This process helped our models to learn faster and required less computer memory.

\subsubsection{Features Aggregation}
We computed a large set of features as described in the earlier section and it was necessary  to reduce the feature space. The aggregation process is depicted in \cref{fig:split1}. The output of the feature extraction process can be described by the matrix:

\begin{equation}
    {A}{_{n}}\:\in \:\mathbb{{R}}^{({S}{_{i}}\times F\times E)_{n}}
\end{equation}
where $n \in N$ and $N$ is the number of the recordings in the data set. $S_{i}$ is the analyzed 8 second cropped segment where $i \in I=\{1,2...,100\}$, $E$ is the number of the selected EEG channels and F is the feature vector having dimension $1\times96$.

The resultant dimension space for $A_n$ is $2717\times21\times100\times96$, as shown in \cref{fig:split1}. We computed the median of all segments as it has been proven to be a successful aggregation function~\cite{Gemein2020}. However, to keep the pattern of the signals and not to lose much information of all analyzed segments, we took the median of the first and last 50 segments separately. Later, we concatenated both the medians and flattened the feature vector space in two-dimensional space to $2717\times4032$.

\subsection{Performance Evaluations}\label{Evaluation}
We report our final results on the held back evaluation set as presented in \cref{tab:TUH abnormal dataset}. The following statistical measures are used to evaluate the performance of the proposed model:
\begin{enumerate}
    \item True-Positive (TP): how often the classifier correctly identifies an abnormal  data sample when it is abnormal.  
    \item True-Negative (TN): how often the classifier correctly identifies a normal data sample when it is normal. 
    \item False-Positive (FP): how often the classifier incorrectly identifies a normal data sample as abnormal.
    \item False-Negative (FN): how often the classifier incorrectly identifies an abnormal data sample as normal. 
\end{enumerate}

The Sensitivity, Specificity and Accuracy of our technique are calculated as:
\begin{equation}
    Sensitivity = \frac{TP}{(TP+FN)}\times 100\%
\end{equation}
\begin{equation}
    Specificity = \frac{TN}{(TN+FP)}\times 100\%
\end{equation}
\begin{equation}
    Accuracy = \frac{(TP+TN)}{(TP+TN+FP+FN)}\times 100\%
\end{equation}

\section{Results and Discussion}
\label{sec3}

\begin{table*}
\resizebox{\linewidth}{!}{
\begin{threeparttable}

\caption{Comparison of performance with other systems.}
\label{tab:Comparison_ofs}

\begin{tabular}{|l|l|l|l|l|l|l|}
\hline
                                          & Study  & ACC(\%) & Sens(\%) & Spec(\%) & Architecture & Features \\ \hline
\multirow{2}{*}{Handcrafted Features}    & Lopez de Diego\cite{Nannen:Thesis:2017}   & 78.8    & 75.4     & 81.9 & CNN+MLP & Cepstral coefficients   \\ \cline{2-7} 
                                          & Lukas et al.\cite{Gemein2020}    & 85.9    & 77.8     & 92.7 & RG+SVM & DFT CWT DWT HT   \\ \hline
\multirow{6}{*}{Deep Learning models} & Yildirim et al.\cite{yildirim2018deep}      & 79.34   &          &      &1D-CNN  &  \\ \cline{2-6} 
                                          & Schirrmeister et al.\cite{Schirrmeister2018}    & 85.4    & 75.1     & 94.1 & BD-Deep4 &    \\ \cline{2-6} 
                                          & Roy et al.\cite{roy2019chrononet}& 86.57   &          &     & ChronoNet    & \\ \cline{2-6} 
                                          &Amin et al.\cite{Amin2019}\tnote{*} & 87.32   & 77.78    & 94.76    & ALexNet+SVM&\\ \cline{2-6} 
                                          & Alhussein et al.~\cite{Alhussein2019}\tnote{*}& 89.13  & 78.78    & 94.76   & AlexNet+MLP& \\ \cline{2-6}
                                          & Lukas et al.\cite{Gemein2020}  & 86.2      & 79.7       & 91.6      & BD-TCN &\\ \hline
\multirow{3}{*}{Proposed models}               & \textbf{CatBoost classifier}     &     \textbf{87.68}    &    \textbf{83.3}      &   91.33    &GBDT & WPD    \\ \cline{2-7} 
                                          & XGBoost classifier &   86.59      &  \textbf{80.9}      & 91.33   & GBDT & WPD     \\ \cline{2-7} 
                                          & LightGBM classifier   &    86.59    &     \textbf{81.7}     & 90.66     & GBDT & WPD    \\ \hline
    
\end{tabular}
\begin{tablenotes}                                      
    \item[*]extra closed source data is used for training.    
\end{tablenotes}
\end{threeparttable}}
\end{table*}
EEG signals from TUH EEG abnormal database~\cite{Nannen:Thesis:2017} are decomposed into sub-bands using WPD, \cref{fig:overall} shows the overall architecture. Wavelet function db4 up to eight levels is used to get the detail and approximation coefficients; the selected coefficients are: [D1, ... , DDDDDDDD8, A1, ... , AAAAAAAA8] (represented by black borders in \cref{fig:newFeatures}). For each level of decomposed coefficients, we computed the MAV, AVP, SD, RMAV, SKEW and KURT. After normalization, the median function is used as an aggregation function which reduces the size of the feature matrix as shown in \cref{fig:split1}. The final feature matrix is later used as an input into three different GBDT classifiers: XGBoost, LightGBM and CatBoost.

\begin{figure*}
	\centering
	\includegraphics[ width=\linewidth]{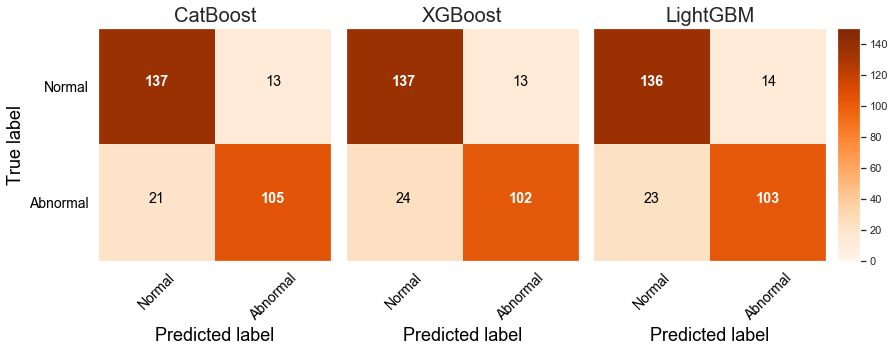}
	\caption {Confusion matrices of three classifiers over independent final evaluation set.}
	\label{fig:cMatrix}
\end{figure*}

\cref{fig:cMatrix} presents the confusion matrices for proposed classifiers, and ~\Cref{tab:Comparison_ofs} presents a comparison of the performance of our models with different state-of-the-art techniques using the same dataset. The confusion matrices indicate that all models achieved higher scores on true-positives resulting in achieving high sensitivity which is very crucial in the field of medicine, especially when the diagnosis is automated~\cite{golmohammadi2019automatic}.

The best result for our model was achieved by using CatBoost classifier. The accuracy, sensitivity and specificity achieved are 87.68\%. 83.3\% and 91.33\%, respectively. This result is the highest as compared to the other state-of-the-art techniques that were trained on the same dataset. Alhussein et al.~\cite{Alhussein2019} have reported slightly higher accuracy than our proposed models; however, it is mentioned in their study that they have pre-trained their model on extra private normal EEG dataset. The details of the extra dataset are not provided, and it is not an open-source dataset and not available to researchers. Compared to other published studies that use the same data, Roy et al.~\cite{roy2019chrononet} reported the accuracy of 86.57\%. Thus our proposed model improves the accuracy by more than 1\% using CatBoost classifier. 

Overall, our proposed models achieve the best sensitivity results for all three classifiers as compared to the existing state-of-the-art techniques while maintaining competitive results for specificity. This shows that our proposed feature extraction is highly optimal and plays a critical role in creating a feature space which can easily be classified using good classifiers.

Furthermore, comparing the results of our models with the other state-of-the-art feature engineering techniques~\cite{Gemein2020,Lopez}, we found that our models with all three classifiers perform better and achieved accuracies of 87.68\%, 86.59\% and 86.59\% for CatBoost, XGBoost and LightGBM classifiers respectively. This shows that our feature engineering technique extracts critical features which can help the classifier to classify the data easily.

\begin{figure}
	\centering
	\includegraphics[ width=\linewidth]{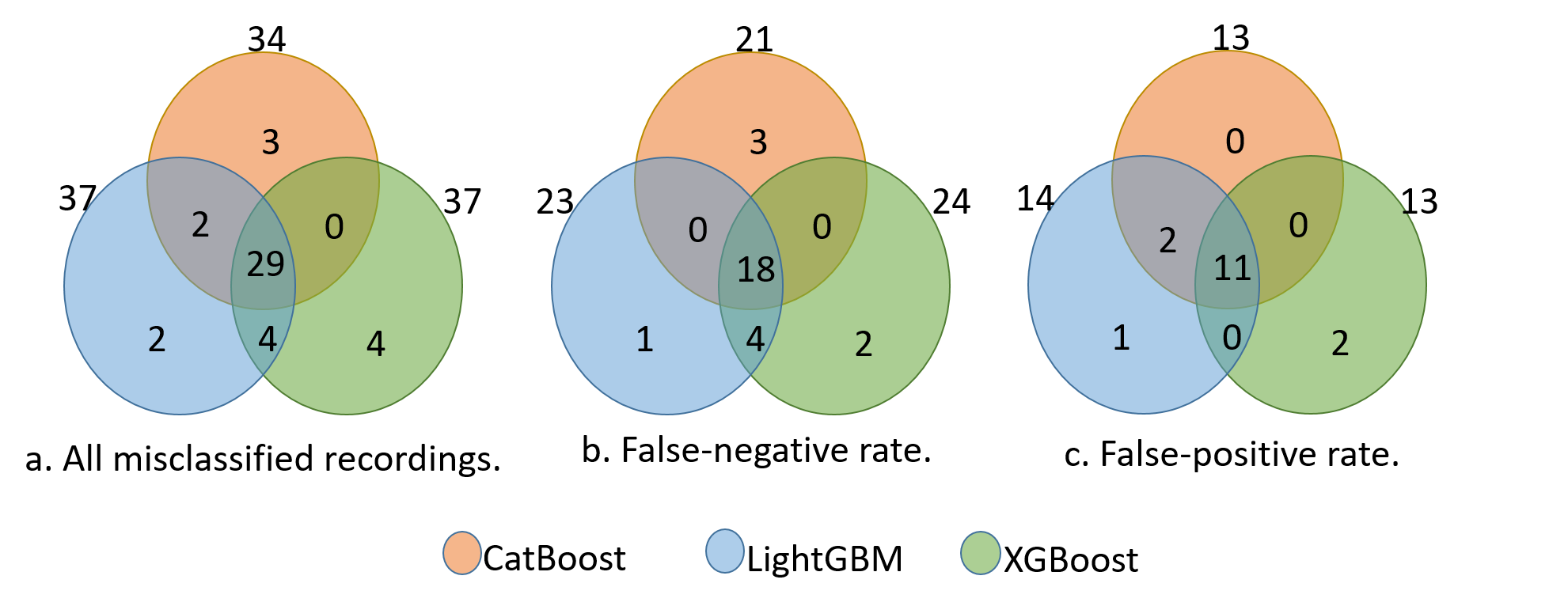}
	\caption {Venn diagrams for misclassification recordings by different classifiers.}
	\label{fig:missclassified1}
\end{figure}

We analyzed the misclassified recordings of our three classifiers as shown in the Venn diagrams in \cref{fig:missclassified1}. We observed in Figure \ref{fig:missclassified1}a that all classifiers misclassified the same 29 recordings. The individual misclassifications range up to four cases only, while common misclassifications of two recordings are observed for CatBoost and LightGBM classifiers and four recordings for LightGBM and XGBoost classifiers. Similarly, we observed from the false-negative Venn diagram for our models \cref{fig:missclassified1}b that all models misclassified 18 recordings as normal, while there were four common misclassifications between LightGBM and XGBoost. The individual misclassifications of one recording, two recordings and three recordings for LightGBM, CatBoost, and XGBoost classifiers respectively are found as depicted in \cref{fig:missclassified1}b. Moreover, we also investigated the false-positive rates as shown in \cref{fig:missclassified1}c. We found that eleven recordings were wrongly classified as abnormal by all classifiers, while CatBoost and LightGBM mutually misclassified two recordings and there was one common misclassification for LightGBM and XGBoost. The individual misclassification is one recording only for the  XGBoost classifier. Overall, it can be observed that more misclassifications are attributed to False-negatives which can be related to the higher number of normal recordings in the dataset.

The analysis of results presented in Figures~\ref{fig:missclassified1}a,~\ref{fig:missclassified1}b, and \ref{fig:missclassified1}c shows that there are many common misclassifications by all classifiers, which can be due to the reason that the extracted features may need further improvement. We can assume that the characteristics of these misclassifications are due to the fact that we took the median of the features of all segments (as presented in \cref{fig:split1}) to reduce the dimensions of the feature space. In addition to the median function, the inter-rater agreement of TUH Abnormal EEG Corpus is higher than the number typically reported in the literature which may result in low performance of CAD system as pointed out by Lukas et al.~\cite{Gemein2020}. The results of this study are consistent with published researches in this context, with a higher rate of specificity rather than sensitivity. The results of this study also agree with findings of Lucas et al.~\cite{Gemein2020}  that the accuracy results for the abnormal EEG detection are within a range of 81 to 86\% as it is the case with LightGBM and XGBoost; however, we push this limit by more than 1\% by the CatBoost classifier. It is also important to remember that in our current study, we only utilized 13 minutes per recording. It is possible that some pathological evidence exists in the remaining parts of recordings.

\section{Conclusion}\label{sec4}
EEG is rich in information about brain health status and is regularly used to decide whether the brain activity is healthy or not. Manual examination of EGG is cumbersome, so the automation of this process can save time for physicians and perhaps support the final diagnostic decision. EEG binary classification is used to isolate the abnormal signals from normal ones so that the mental condition related to the abnormality can be investigated further.

We proposed a novel technique which decomposes EEG signals into time-frequency sub-bands through WPD feature extraction and selects a set of features in a reduced aggregated feature space. Three different classifiers based on GBDT are used to automatically detect abnormal brain signals in multi-channel EEG recordings using publicly available data TUH Abnormal EEG Corpus. Experimental results demonstrate that the proposed technique performs better as compared to the existing state-of-the-art techniques using the same dataset. We evaluated the performance of different classifiers and found the accuracies to be 87.68\%, 86.59\%, and 86.59\% for CatBoost, XGBoost, and LightGBM classifiers, respectively. Together these results provide important insights into the usefulness of proposed feature selection technique for automatic detection of abnormal brain signals in multi-channel EEG recordings and can serve as a promising choice for medical application in the future. In future, we plan to use the proposed technique to classify multiple brain-related disorders.

\section{Consent for publication}
All authors have read and approved the manuscript.

\section{Declaration of Competing Interest}
The authors confirm that there are no known conflicts of interest associated with this work.

\end{document}